\documentclass{osa-article}

\journal{osajournal}

\articletype{Research Article}

\begin{document}

\title{A scanning planar Yagi-Uda antenna for fluorescence detection}

\author{Navid Soltani,\authormark{1,6,*} Elham Rabbany Esfahany,\authormark{1,6} Sergey  I. Druzhinin,\authormark{2,6} Gregor Schulte,\authormark{2,6} Julian M\"{u}ller,\authormark{3,6} Florian Sledz,\authormark{1,6} Assegid Mengistu Flatae,\authormark{1,6} Benjamin Butz,\authormark{3,6} Holger Sch\"{o}nherr,\authormark{2,6} Nemanja Marke\v{s}evi\'{c} \authormark{1,5} and Mario Agio \authormark{1,4,6}}

\address{\authormark{1}Laboratory of Nano-Optics, University of Siegen, Siegen 57072, Germany\\
\authormark{2}Physical Chemistry I, University of Siegen, Siegen 57076, Germany\\
\authormark{3}Micro- and Nanoanalytics Group, University of Siegen, 57076 Siegen, Germany\\
\authormark{4}National Institute of Optics (INO), National Research Council (CNR), Florence 50125, Italy\\
\authormark{5}Currently with Nanoscience Center, University of Jyväskylä, Jyväskylä 40014, Finland\\
\authormark{6}Research Center for Micro and Nano-Chemistry and Engineering (C$\mu$), Siegen 57076, Germany}

\email{\authormark{*}navid.soltani@uni-siegen.de, mario.agio@uni-siegen.de} 



\begin{abstract}
An effective approach to improve the detection efficiency of nanoscale light sources relies on a planar antenna configuration, which beams the emitted light into a narrow cone. Planar antennas operate like optical Yagi-Uda antennas, where reflector and director elements are made of metal films.
Here we introduce and investigate, both theoretically and experimentally, a scanning implementation of a planar antenna. Using a small ensemble of molecules contained in fluorescent nanobeads placed in the antenna, we independently address the intensity, the radiation pattern and the decay rate as a function of distance between a flat-tip scanning gold wire (reflector) and a thin gold film coated on a glass coverslip (director).
The scanning planar antenna changes the radiation pattern of a single fluorescent bead and it beams light into a narrow cone down to angles of 45$^{\circ}$ (full width at half maximum). Moreover, the collected signal compared to the case of a glass coverslip is larger than a factor of 3,  which is mainly due to the  excitation enhancement. These results offer a better understanding of the modification of light-matter interaction by planar antennas and they hold promise for applications, such as sensing, imaging and diagnostics.
\end{abstract}

\section{Introduction}

Despite advances in the fields of optics and nano-optics, the detection of fluorescent molecules remains a big challenge. In general, low excitation cross-sections at room temperature, low quantum yields of molecules and their dipolar emission patterns are only some of the problems that we have to face to detect low fluorescence signals.
To overcome these issues a lot of different techniques have been developed over the years, ranging from simple ones such as the use of high numerical aperture (NA) mirrors and objectives~\cite{drechsler2001confocal,enderlein2000theoretical} to the complicated fabrication techniques for  creating nanostructures that can change the optical properties of the emitters and facilitate their detection, such as  optical microcavities~\cite{kaupp2016purcell,huckabay2011whispering}, photonic nano-wires~\cite{claudon2010highly,babinec2010diamond} and nano-antennas~\cite{kinkhabwala2009large, langguth2017nano}. 
For example, in the vicinity of plasmonic nanoparticles illuminated by laser light,  the excitation intensity can increase by orders of magnitude~\cite{russell2012large, sugimoto2018hybridized}. In addition, plasmonic particles spectrally matched with fluorescent molecules can also increase the emission rate as well as the quantum efficiency, making them  significantly brighter~\cite{bidault2016picosecond, flauraud2017plane, acuna2012fluorescence,lu2020quantum}.  It is also well known that some of these structures operate like antennas to change the radiation pattern of emitters and make them  radiate into smaller angles \cite{aouani2011bright,akselrod2014probing, xia2021enhanced, kuhn2008modification}.
For example, a single quantum dot coupled to a Yagi-Uda  nano-antenna shows strongly polarized and highly directional luminescence~\cite{curto2010unidirectional}. However, the production of these nano-objects demands complicated fabrication techniques and precise positioning of the emitters in the antenna's hotspot. 


Recently, a high collection efficiency up to 99\% from a single molecule was demonstrated with dielectric \cite{lee2011planar}  and metallo-dielectric \cite{chu2014experimental} planar antennas. Since a planar antenna has a broadband resonance, the position of the emitter is not crucial and the fabrication of planar structures is relatively simple and straightforward. However, these antennas direct the radiation towards the microscope objective at wide angles, which requires a high-NA  objective for efficient collection. 

A planar Yagi-Uda antenna, which consists of two thin metallic films, a reflector and a director, not only beams the radiation pattern into smaller angles, but also enhances the emission rate of a molecule \cite{checcucci2017beaming}. In this system, the distance from the emitter to the reflector is not a sensitive parameter and it can vary between $\lambda / 6 n$ to $\lambda / 4 n$ where, $n$ is the refractive index of the medium between them. This type of antenna can also enhance the collection efficiency for light sources embedded in high-refracting index materials~\cite{galal2017highly,galal2019highly}.

So far, planar Yagi-Uda antennas have been designed to operate in a fixed configuration and the emitters are embedded in a solid matrix. 
However, tunable microcavities or antennas allow more operational freedom and they can change the optical properties of emitters in a controlled manner~\cite{miguel2013cavity, chu2014experimental,kelkar2015sensing,wang2019turning, casabone2018cavity}, hence offering additional possibilities and paving the way towards potential applications.
In this work, we introduce and investigate a scanning implementation of a planar Yagi-Uda antenna, which could even be exploited for fiber collection~\cite{soltani2019planar}. Our main goal is to examine how the antenna operates as a function of the reflector-director distance and tune for instance the collection efficiency, the quantum yield and the radiation pattern of a nanoscale light source. 
This approach can be applied to different kinds of nanoscale light sources, such as molecules, quantum dots, beads, and it can be operated in different environments, as in water or at cryogenic temperatures. Therefore, it represents a novel platform for enhanced spectroscopy and sensing.


\section{Layout of the problem and simulation results}

Although previous works simulate the radiation pattern and the maximum collection efficiency for the optimal distance between the emitter and the antenna elements~\cite{checcucci2017beaming,galal2017highly}, we extend this approach and simulate the parameters for continuous reflector-director distances even larger than  2\,$\lambda$ (up to 1.5 $\mu$m). Moreover, in a realistic simulation of fluorescence detection, one should also consider the excitation enhancement and the quantum efficiency of the molecule ($\eta_0$). Therefore, we examine the  collection efficiency of a dipole in a scanning planar antenna excited by a plane monochromatic electromagnetic wave.

\begin{figure}
  \includegraphics[width=12cm]{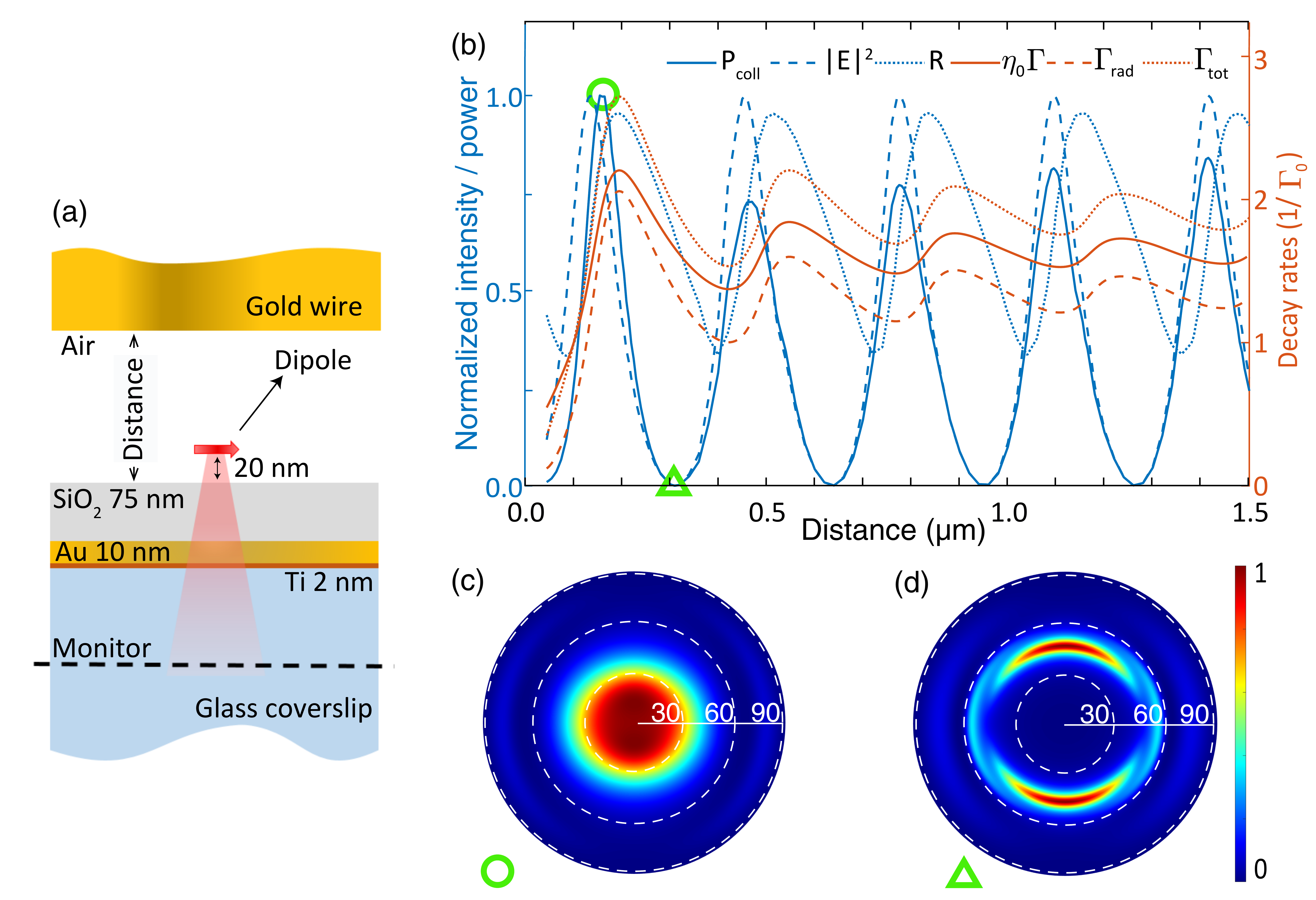}
  \centering
  \caption{\textbf{Hertzian dipole in a scanning planar antenna}. (a) Simulation layout of a dipole excited  with a 636 nm monocromatic light and its emission is at 680 nm in the antenna configuration. The glass coverslip and the gold wire (reflector) are semi-infinite. The values of the fixed parameters are given in the figure. The distance between the SiO$_2$ layer and the gold reflector (air gap) varies from 45 nm to 1500 nm.  The reflectivity $R$ and the excitation intensity $|E|^2$  are recorded with a frequency-domain monitor. (b) Plane-wave intensity at the dipole position, $|E|^2$, reflectance $R$ and collection efficiency $P_\mathrm{coll}$ plotted as a function of the air gap. 
  Modification of the decay rate $\Gamma$ (assuming the quantum yield of 0.7) as well as $\Gamma_\mathrm{tot}$ and $\Gamma_\mathrm{rad}$ also plotted as a function of the air gap. The far-field radiation patterns, corresponding to the  maximum and minimum of $P_\mathrm{coll}$, are highlighted with a circle and a triangle and they are represented in (c) and (d), respectively. The FWHM of the intensity profile in (c) is 45$^{\circ}$.}
  \label{theory_emission}
\end{figure}

We take into consideration only horizontal projections of the dipole moments of emitters, since the vertical ones are strongly coupled to the surface plasmon polariton modes of the metal films and they do not contribute to the emission in the far field~\cite{checcucci2017beaming}. Thus, for emitters with random orientation, the coupling efficiency would scale by a factor of 2/3~\cite{soltani2019planar}. 
In our calculations, the gold layer, which represents the director is 10~nm thick and the reflector is a gold mirror movable along the $z$-axis. As already demonstrated, the optimal distance between the dipole and the director should be approximately between $\lambda/6n$ and $\lambda/4n$~\cite{checcucci2017beaming}.
For this reason, we need a spacer to provide such a distance in a realistic simulation. The ideal spacer should be optically transparent and the emitters should not be quenched in its vicinity~\cite{rodriguez2016self}.
Therefore, the dipole is located 20~nm above a 75~nm thick SiO$_2$ layer used to separate the dipole from the director (Fig.~\ref{theory_emission}a). This distance is chosen to mimic the 20~nm radius of fluorescent beads on the SiO$_2$ layer, which we model as a single dipole.
The parameters are calculated  in the range of distances from 45 nm to 1500 nm between a semi-infinite gold mirror (reflector) and the SiO$_2$ layer.  To simulate the realistic structure we also consider that the director layer is on the top of a 2~nm titanium (Ti) film, which is supported by a semi-infinite borosilicate glass (Fig.~\ref{theory_emission}a).

We calculate the normalized total decay rate $\Gamma_\mathrm{tot}/ \Gamma_0$ and the normalized radiative decay rate to the far field $\Gamma_\mathrm{rad} / \Gamma_0$ by respectively considering the power dissipated and radiated by a Hertzian dipole emitting at 680 nm in the antenna configuration with respect to free space~\cite{taminiau2008enhanced,kaminski2007finite}. 
Here, $\Gamma_0$ represents the radiative decay rate of the dipole in vacuum. Hence, $\Gamma_\mathrm{tot} / \Gamma_0$ corresponds to the Purcell factor for an emitter with an intrinsic quantum yield $\eta_0 = 1$ and it reaches a value of about 2.8 at the first maximum (Fig.~\ref{theory_emission}b). A normalization with respect to a dipole near a glass coverslip would reduce the Purcell factor by a factor of about 1.25 (average refractive index between SiO$_2$ and air), since the radiative decay rate is directly proportional to the refractive index. 
We also consider the dipole excitation at a wavelength of 636 nm, which for simplicity is assumed to be a plane-wave. To estimate the enhancement of the excitation rate we calculate the electric-field intensity $|E|^2$ at the dipole position. The plane-wave reflectivity $R$ is also calculated to relate the excitation enhancement with this measurable quantity. We numerically solve  the electromagnetic problem (computation of $|E|^2$ and $R$) using FDTD Solutions (Lumerical)~\cite{lumerical} and semi-analytically (computation of the decay rates and of the radiation patterns)~\cite{neyts1998simulation}.

Figure~\ref{theory_emission}b plots $|E|^2$ (dashed blue curve), normalized to its maximum value, and $R$ (dotted blue curve), normalized to the incident power, as a function of the air gap in the planar antenna. Since we use a plane wave, both quantities oscillate with a periodicity of $\lambda/2$, where $\lambda=636$ nm, and with a constant maximum height, like in a Fabry-Perot etalon. Notice that the maxima and minima of $|E|^2$ and $R$ do not coincide, because the planar antenna is a kind of asymmetric cavity and $|E|^2$ is calculated at the fixed dipole position, which is not at the maximum of the cavity modes.

Since we consider a detection with a high-NA objective, the collected power $P_\mathrm{coll}$ can be assumed proportional to the product of $|E|^2$ with $\Gamma_\mathrm{rad}$, as we will demonstrate later, and to be nearly independent of the radiation pattern. In Fig.~\ref{theory_emission}b, $P_\mathrm{coll}$ (solid blue curve) is plotted for different values of the air gap between the reflector and the SiO$_2$ layer.
$P_\mathrm{coll}$ oscillates with $|E|^2$ and $\Gamma_\mathrm{rad}$ and therefore its behaviour is determined by both excitation and radiation enhancement and by the radiation pattern. 

When $R$ has a minimum, $P_\mathrm{coll}$ is not maximal. It is also remarkable to note that even when $|E|^2$ is maximal, $P_\mathrm{coll}$ is not maximal for the first peak. This indicates the different contributions of $|E|^2$ and  $\Gamma_\mathrm{rad}$ to the emission. Nonetheless, when the distance increases $\Gamma_\mathrm{rad}$ has a smaller contribution and $P_\mathrm{coll}$ oscillates with $|E|^2$.

To understand the role of the decay rates, Fig.~\ref{theory_emission}b plots $\Gamma_\mathrm{rad}$ (dashed red curve), $\Gamma_\mathrm{tot}$ (dotted red curve) and the inverse excited-state lifetime $\Gamma$ (solid red curve), calculated by $(1-\eta_0) \Gamma_0/ \eta_0 + \Gamma_\mathrm{tot}$ for an intrinsic quantum yield $\eta_0=0.7$~\cite{chance1978molecular}, which is close to that of our fluorescent beads in the experiment.

Since the beads have an intrinsic quantum yield of 0.7, the modification of the excited state decay rate is given by  $ \eta_0  \Gamma / \Gamma_0$. This value oscillates around 2 as shown in Fig.~\ref{theory_emission}b because the normalization is with respect to a dipole in free space.

Because part of the emitted light is absorbed by the metallic elements of the antenna, it is relevant to pay attention to the antenna efficiency $\eta_\mathrm{a}=\Gamma_\mathrm{rad}/\Gamma_\mathrm{tot}$, which represents the fraction of photons reaching the far field.
This efficiency is about 0.6 and it is almost independent of the distance between the reflector and the director (not shown). Moreover, $\Gamma_\mathrm{rad}$ is the quantity that contributes to  $P_\mathrm{coll}$ as we show here. 
In fact, below saturation the fluorescence signal $S$ is proportional to the product of intensity, quantum yield and decay rate, i.e.  $S=|E|^2\eta\Gamma$, where $\eta=\eta_0/[(1-\eta_0)\Gamma_0/\Gamma_\mathrm{rad}+\eta_0/\eta_\mathrm{a}]$ and $\Gamma_0$ is the radiative decay rate without antenna. Since $\Gamma=(1-\eta_0) \Gamma_0 / \eta_0 + \Gamma_\mathrm{tot}$, it is easy to show that $S=|E|^2\Gamma_\mathrm{rad}$.
Now, considering the fact that the high-NA objective does not significantly modify the collection efficiency, we can assume $P_\mathrm{coll} \simeq S$.

The far field emission patterns of the dipole in the first local maximum and minimum of $P_\mathrm{coll}$ are shown in Fig.~\ref{theory_emission}c and d, respectively. In Fig.~\ref{theory_emission}c, the radiated intensity is peaked at the center and its full width at half maximum (FWHM) is 45$^{\circ}$. On the other hand, in Fig.~\ref{theory_emission}d, the radiated intensity at small angles drops almost to zero.

\section{Experimental methods}

\subsection{Antenna fabrication and setup configuration}
To verify our theoretical predictions, we design a planar antenna according to the parameters used for the calculations in the previous section.
For light sources, we choose bright fluorescent beads with a 40 nm diameter (ThermoFisher, 0.04 $\mu$m, dark red fluorescent 660/680), highly concentrated fluorescent molecules embedded in polystyrene (approximately 350 molecules), which are less prone to bleaching. 
Since our goal is to address individual beads, we dilute them in pure water (MilliQ 18.2 M$\Omega \cdot$cm resistivity) and subsequently spin-cast them on a substrate in order to obtain spatially separated beads (at least few $\mu$m mutual distance). 

\begin{figure}[h]
\centering
\includegraphics[width=12cm]{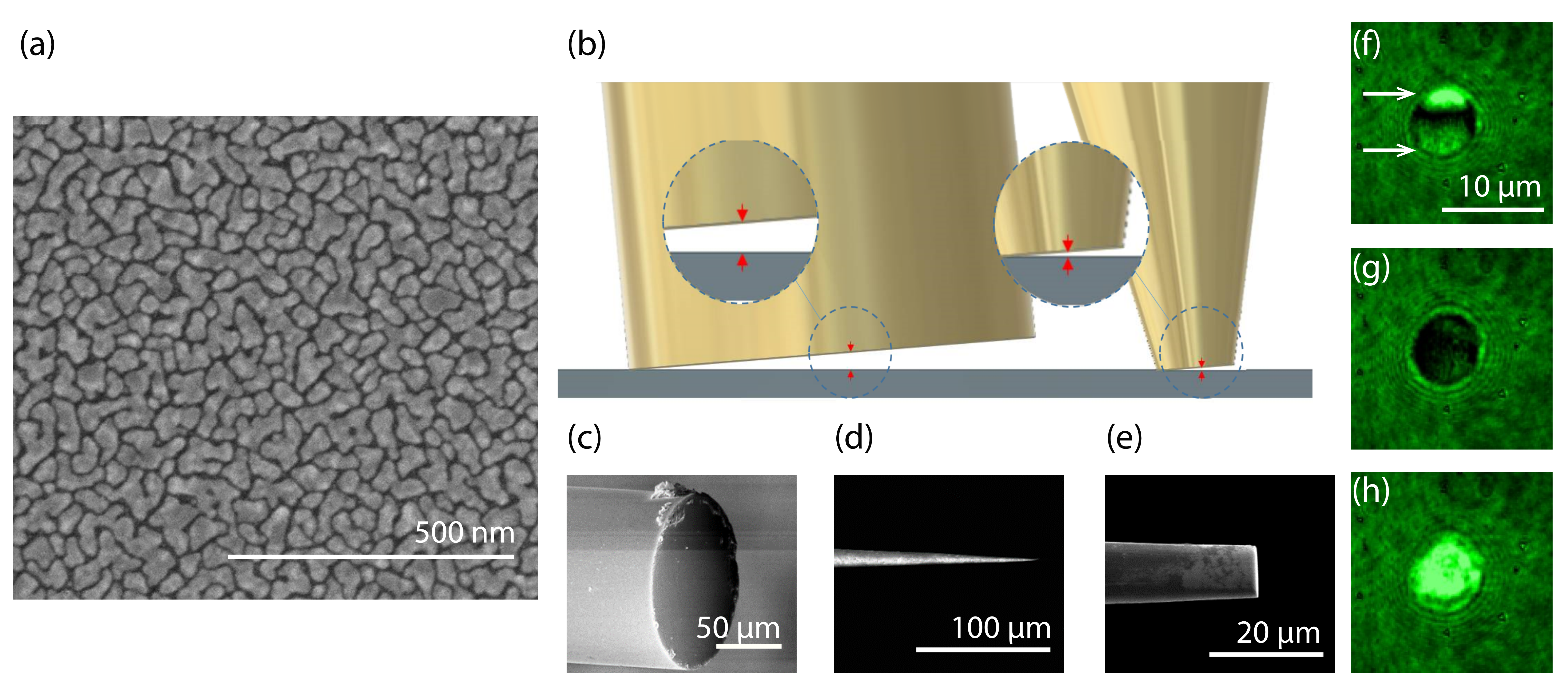}
\caption{\textbf{Fabrication of gold director and wire reflector}. (a) Scanning electron microscopy (SEM) image of a 10~nm gold layer on glass coverslip. Instead of flat films,  ultrathin islands are formed. The surface roughness of the film (director) is 10.2 nm peak to peak, which was measured by atomic force microscope (data not shown).
(b) Schematic representation of a tilted gold wire before and after the thinning process. The distance from the substrate to the center of the wire varies significantly depending on the wire's diameter. The insets are magnified images of the central part the gold wire tip and the underneath substrate. SEM image of a gold wire (c) before and (d) after electrochemical etching. (e) SEM image of the wire tip after FIB cutting. (f) The interference pattern formed by a tilted wire and the gold director. Since only two maxima are visible (white arrows), the estimated angle between  the wire and the substrate is around $2.6^{\circ}$. In (g) and (h) this angle is close to zero and thus these images represent consecutive dark and bright fringes due to 316~nm wire displacement.}
\label{etching1}
\end{figure}

A planar antenna consists of two main parts, a director and a reflector. To produce the director, a glass coverslip (PLANO, L42342 St\"arke \#1) is firstly coated with 2 nm titanium (Ti) \cite{werner2009optical} which plays the role of an adhesion layer between the glass coverslip and the 10~nm gold film \cite{olmon2012optical} (director). Furthermore, we evaporated 75 nm SiO$_2$ layer as a spacer on top of Au film (similar to the structure presented in Fig.~\ref{theory_emission}a). All the material deposition processes are preformed by e-beam evaporation (Edwards E306A) with an evaporation rate of 0.1 nm/s.  As shown in Fig.~\ref{etching1}a, the 10~nm gold on layer on Ti does not form a uniform film, but ultrathin islands.
In these kind of structures, polarized light can excite localized surface plasmons~\cite{sun2012temperature}.
 
The numerical calculations indicate that the air gap between the SiO$_2$ layer and the gold mirror should be 160 nm to receive the maximal directional emission from a Hertzian dipole (Fig.~\ref{theory_emission}b). However, we note that a small tilt angle between reflector and substrate (Fig.~\ref{etching1}b), can make the center gap larger than the optimal distance. One way to mitigate this issue is to reduce the diameter of the gold mirror, e.g. by etching the tip of the wire. From a simple geometrical consideration we can easily demonstrate that even for 3$^{\circ}$ tilt angle between the wire and the substrate, the distance between the centre of the gold wire and the substrate is 183~nm for a 7~$\mu$m wire and 5.23~$\mu$m for a 200~$\mu$m wire diameter.  

In order to obtain gold tips with a 5-10 $\mu$m diameter, we follow the procedure of Ref.~\cite{ren2004preparation}. Namely, we first vertically submerge a 200~$\mu$m thick gold wire (Fig.~\ref{etching1}c) up to 1 cm into an electrolyte solution (37 \% hydrochloric acid (HCl)). By applying a 3-Volt potential difference between the anode, the gold wire, and a platinum cathode, the gold wire is etched down to a cone shaped micro-tip (Fig.~\ref{etching1}d). Several parameters such as the time of applying the voltage, the length of the wire inside the HCl and the HCl concentration, play a role in the fine etching procedure. However, since we only need a sharp tip with a radius of a few micrometers these parameters are not crucial. 
After the etching process in order to obtain a flat mirror, the tip of the wire is subsequently cut by Focused Ion Beam (FIB) milling (FEI Helios Nanolab 600) as shown in Fig.~\ref{etching1}e. The face diameter of the top is determined by the position of the FIB cut. To achieve a flat tip and to avoid edge rounding at the tail of the ion beam, FIB milling is performed at 30 kV and with the smallest aperture possible while maintaining reasonable cutting times. Depending on the diameter of the wire tip, the used apertures typically resulted in a beam current of 0.92~nA to 2.8~nA. As FIB milling is well-suited for high resolution electron microscopy sample preparation, the local roughness of the tip is expected to be within a few nm.

The contact point and the tilt angle can be measured by monitoring interference fringes as shown in Fig.~\ref{etching1}f-h. For a 7-$\mu$m gold tip and 636 nm coherent light, the minimal angle to obtain only two bright fringes is 2.6$^{\circ}$ (Fig.~\ref{etching1}f). In the experiment this angle can be adjusted by a manual tilting stage (Newport M-561-TILT) below this value to overcome the central gap distance (Fig.~\ref{etching1}g,h).

\subsection{Optical setup}
The optical setup is based on an epi-fluorescence inverted microscope (Supporting Information Fig.~S1a). A pulsed laser with a wavelength of 636 nm and a repetition rate of 40 MHz (Picoquant, LDH-P-C-640B) is coupled to a commercial inverted  microscope (Zeiss, Axio Observer 3) and the laser light is focused with an oil objective (Zeiss, Plan-Apochromat 63x/1.4) onto the sample. A motorized stage (ASI, PZ-2000FT) controls the position of the glass substrate with the spin-casted beads, in the $xyz$-directions in order to address a single bead.
The reflector is aligned above the focal point of the objective with a home-built module on the top part of the microscope. A piezo controller (Newport, NPM140SG) moves the wire in the $z$-direction in order to set the distance between the reflector and the bead, positioned in the focal spot of the laser. The piezo-controller movement is  monitored by a closed loop piezo stack amplifier (Newport NPC120SG).

The fluorescence signal from a single bead is collected by the microscope objective and simultaneously sent to the CMOS camera (Andor, Zyla 4.2 Plus) and the single-photon avalanche photodiode (SPAD) (Excelitas, SPCM-AQRH-TR), with a 95:5 Transmitance:Reflectence ratio defined by a beam splitter (BS). A Time-Correlated Single-Photon Counting (TCSPC) device (PicoQuant, Pico Harp 300) measurs the photon count rate. The time resolution in this experiment is less than 250~ps, which is the time  resolution of SPADs. The radiation pattern was obtained by Back-Focal Plane (BFP) imaging with the CMOS camera. 

The setup contains also a laser clean-up filter (CF 637/7) with central wavelength of 637 nm and 7 nm bandwidth, a flippable wide-field lens (WFL) used for monitoring the beads in the alignment process, a dichroic mirror (DM) with a 650-nm cut-off wavelength, a long-pass (LP) and a band-pass (BP) filter (LP 650, BP 680/42, with central wavelength of 680 nm and 42 nm bandwidth), whose position can change based on the experiment (Fig.~S1b). Flip mirrors (FM1 and FM2) can send the light to the CMOS camera, the spectrometer (Ocean Optics QE pro) or to another single-photon avalanche photodiode (SPAD1). A
back-focal plane lens (BFL) is used to image the radiation pattern. The inset of Fig.~S1a sketches the 3D model of gold wire in the antenna configuration. The distance of the wire from glass coverslip is monitored by a tilted objective mounted on the home-built stage  such that the wire tip is clearly visible while approaching the substrate surface. 
For precise alignment, the tilting of the gold wire is observed by the laser interference pattern also collects by the microscope objective and corrects by a manual tilt stage as described in Fig.~\ref{etching1}f-h. Moreover, the contact point between the reflector and the substrate is determined by interference fringes. Namely, once the wire and the substrate are in contact, further attempt to approach does  not result in any visible interference pattern change. We take this position as a starting point of the wire retraction.
 
The experiment is performed at room temperature and the vibration of the antenna system is estimated by interference patterns to around 40~nm. Therefore, in all steps we measure the position several times and average the result. This further helps us to decrease the vertical position error to $\pm$5~nm at each step.

\section{Experimental results and discussion}

\begin{figure}[h]
  \includegraphics[width=12cm]{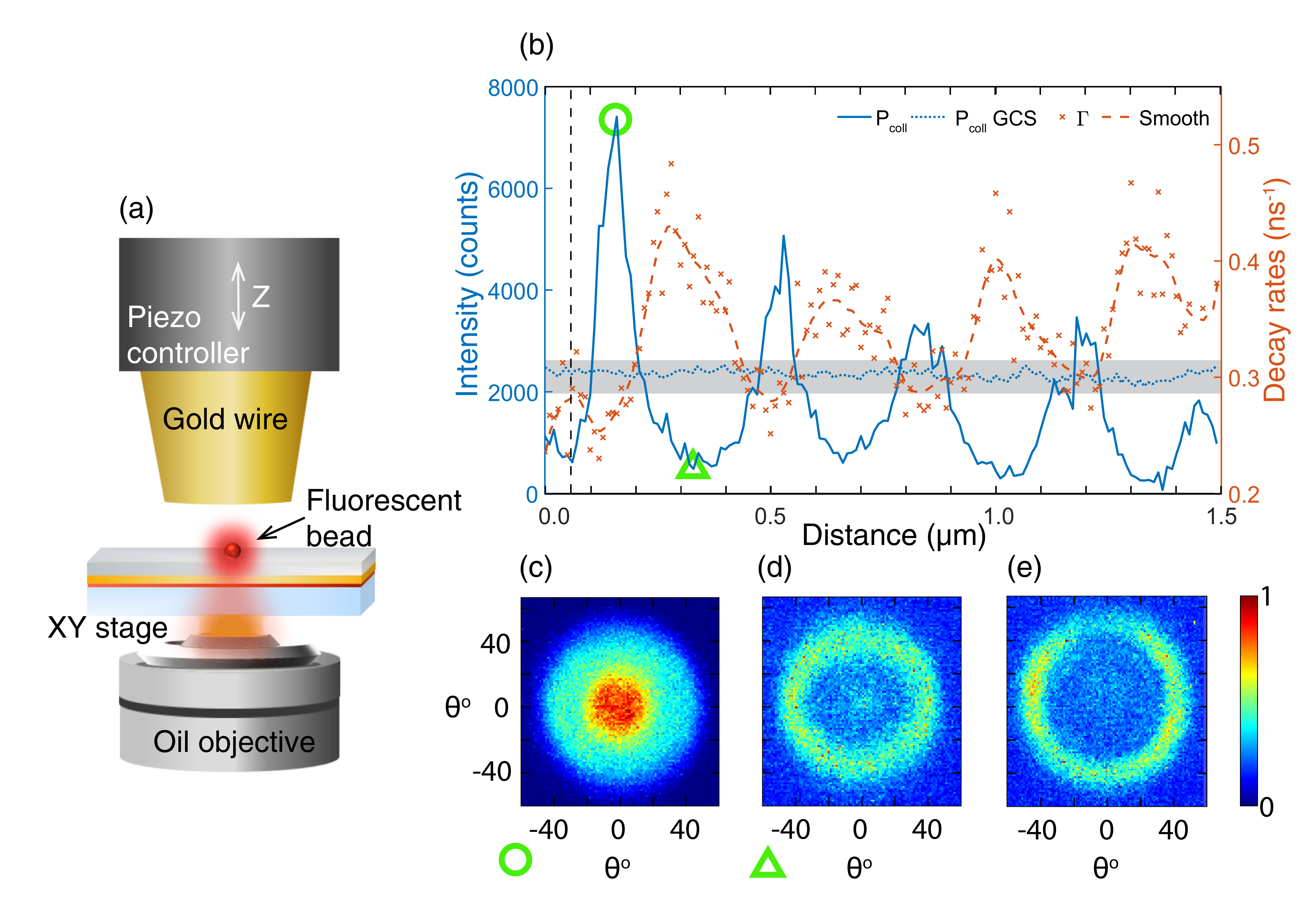}
  \centering
  \caption{\textbf{A fluorescent bead in the scanning antenna configuration}. (a) Schematic representation of the home-built scanning planar antenna setup. The gold wire and the sample can move by piezo controller (along Z-axis) and motorized stage (in XYZ), respectively. (b) Collected fluorescence intensity (counts) and decay rates $ns^{-1}$ measured as a function of the distance between the gold wire and the SiO$_2$ layer. The background intensity is subtracted. The vertical black dashed line indicates the position, where the gold wire and the SiO$_2$ are in contact. The collection intensity of a bead on the glass coverslip is shown with the blue dotted line and the range of this intensity for different beads is showed by the gray shaded area. (c,d) Radiation pattern for two selected positions marked by a circle and a triangle. (e) Radiation pattern of a bead when the reflector is far away. Each BFP image is normalized to its maximum values.}
  \label{Beads_antenna1}
\end{figure}

Here, we investigates the emission pattern, the excited-state decay rate $\Gamma$ and the collected power $P_\mathrm{coll}$ for a single bead in the planar antenna configuration, as a function of the distance between the SiO$_2$ layer and the reflector (air gap), as shown in Fig.~\ref{Beads_antenna1}. The reflector position is adjusted using the piezo controller in a step-wise fashion (10 nm step size), from the contact point with the SiO$_2$ in 150 steps (Fig~\ref{Beads_antenna1}a). At each step the position of the wire and the measured quantities are recorded.
The vertical dashed line in Fig.~\ref{Beads_antenna1}b indicates the contact point. 

In our experiment, the excitation power at the back entrance of the microscope is 3.5 $\mu$W. Since the 10~nm gold director has only 30\% transmittance, the power passing through the director is around 1 $\mu$W.
However, this is sufficient to detect the fluorescent beads and align them in the focus point of the objective using the  $xyz$-stage.

To determine fluorescence lifetimes at each step of the reflector position, we construct fluorescence lifetime decay histograms based on start-stop events of the TCSPC. Using the deconvolution method, we were able to separate fluorescence from background and to reach the single exponential fluorescence decay rate ($\Gamma$) (see Supporting Information for details). 
The integration of the fluorescence decay histogram determines the collected power  ($P_\mathrm{coll}$) as a function of the distance between the reflector and the SiO$_2$ spacer. (Supporting Information Fig.~S2). The variation of $P_\mathrm{coll}$ and $\Gamma$ at different reflector positions are plotted in Fig.~\ref{Beads_antenna1}b  with blue and red curves, respectively.

The decay rate $\Gamma$ in the scanning Yagi-Uda antenna exhibits a periodic behavior with values between 0.48 and 0.27 ns$^{-1}$, i.e. lifetime $\tau$ is between 2.08 and 3.70 ns, and it has a phase shift with respect to $P_\mathrm{coll}$ (Fig.~\ref{Beads_antenna1}b). 
This is due to the fact that the decay rate increases at the cavity resonance for the emission wavelength ($\lambda \simeq 680$ nm), while the detected intensity follows the laser wavelength (excitation enhancement at $\lambda = 636$ nm) and the emission wavelength (enhancement of the radiative decay rate).
The fluctuation of the decay rate, especially at low $P_\mathrm{coll}$ is mainly due to the low intensity of the fluorescence decay from a single bead as well as the lower background caused by  the ultrathin gold islands (director). Although this background is deconvolved from the fluorescence signal of the bead, the signal to noise ratio still remains very low at the low emission intensities (Fig.~S2).  

A smoothed $\Gamma$ (red dashed curve in  Fig.~\ref{Beads_antenna1}b) was obtained by averaging  the decay rate at each point with its 6 neighbours. Moreover, by calculating the ratio between the first maximum and minimum of the smooth curve (1.6) and comparing it to  $\Gamma$ in Fig.~\ref{theory_emission}b, one can estimate the quantum efficiency of the beads, which is  around $\eta_0 = 0.7$~\cite{chance1978molecular,buchler2005measuring}. The modification of the excited-state lifetime due to the Purcell factor of about 2, which is in a good agreement with the simulation results presented in Fig.~\ref{theory_emission}a.

The beads exhibit an emission spectrum peaked at 680~nm with a FWHM of about 38~nm and the excited-state decay rate is about $0.29 \pm 0.05$ ns$^{-1}$ on the glass coverslip (not shown). The collection intensity of a single bead on the glass coverslip (without any additional layer) is shown by the blue dotted line in Fig~\ref{Beads_antenna1}b for reference.
Although the bleaching of the fluorescent molecules inside the bead can potentially reduce its emission, this effect is neglected in our experiments owing to the long bleaching time of the beads (blue dotted line in Fig.~\ref{Beads_antenna1}b). 

The main reason for the reduced $P_\mathrm{coll}$ in the antenna configuration (Fig.~\ref{Beads_antenna1}b) must be attributed to the smaller excitation intensity due to the standing wave pattern formed by the laser. One should take into account that the antenna acts also as a cavity and gives rise to the intensity enhancement (blue dashed line in Fig.~\ref{theory_emission}b ). 

Furthermore, at larger distances, the cavity resonance has a dominant effect and the signal exhibited a series of maxima and minima that follow the excitation wavelength.
Since fluorescent beads contain a different number of fluorescent molecules, the emitted power of the individual beads is not the same. It is therefore difficult to compare the antenna configuration with the glass coverslip in an quantitative manner. However, a comparison of the $P_\mathrm{coll}$ in the antenna with $P_\mathrm{coll}$ for several beads on the glass coverslip, shown by a gray shaded area (Fig.~\ref{Beads_antenna1}b), demonstrates that overall the antenna enhances the fluorescence signal by more than a factor of 3. The dotted curve represents one measurement of $P_\mathrm{coll}$ of a bead on the glass coverslip.

The maximum of $P_\mathrm{coll}$ is at 0.17 $\mu$m distance and the back-focal plane (BFP) image  at this position (Fig.~\ref{Beads_antenna1}c) indicates a highly directional emission.
The FWHM of the emission pattern at the first maximum of $P_\mathrm{coll}$ shrinks to roughly 45$^\circ$. Moreover, the emission pattern at the first minimum of $P_\mathrm{coll}$ is a ring with 70$^\circ$ angle (Fig.~\ref{Beads_antenna1}d), which is also in good agreement with the simulation results shown in  Fig.~\ref{theory_emission}c,d. Since the NA of the objective is high, we can compare the measured intensity with the calculated $P_\mathrm{coll}$ shown in Fig.~\ref{theory_emission}b. The radiation pattern of the bead when the reflector is far away is shown in Fig.~\ref{Beads_antenna1}e and the results are consistent with earlier studies~\cite{zhu2017out,brunstein2018decoding,dasgupta2021direct}.

Inspired by the planar metallo-dielectric antennas ~\cite{chu2014experimental,devilez2010compact} we design a semi-antenna, which is a simplified version of the planar Yagi-Uda antenna (without SiO$_2$, gold director and Ti layers). In Supporting Information (Fig.~S4) we provide the radiation pattern for the single bead influenced by this antenna configuration.  Comparing the radiation patterns for the  planar Yagi-Uda antenna presented above (Fig.~\ref{Beads_antenna1}c) with the semi-antenna, we clearly demonstrate that even a 10~nm thick gold layer (director), with all its imperfections, can dramatically change the radiation pattern (see for comparison Fig.~S4c,d and Fig.~\ref{Beads_antenna1}c,d).

To determine the relationship between $R$ and $P_\mathrm{coll}$, the collected light is split into two channels. We slightly modified the previous setup and, instead of a CMOS camera, we employed SPAD1 for photon detection (Supporting Information Fig.~S1b). In this case, 95\% of the light was passing through a long-pass and a band-pass filter and detected by the SPAD1. Conversely, after being attenuated only by a neutral density (ND) filter (OD=2), the other 5\% of light (laser reflection) was detected by a second photo-detector (SPAD2).

\begin{figure}
  \includegraphics[width=8cm]{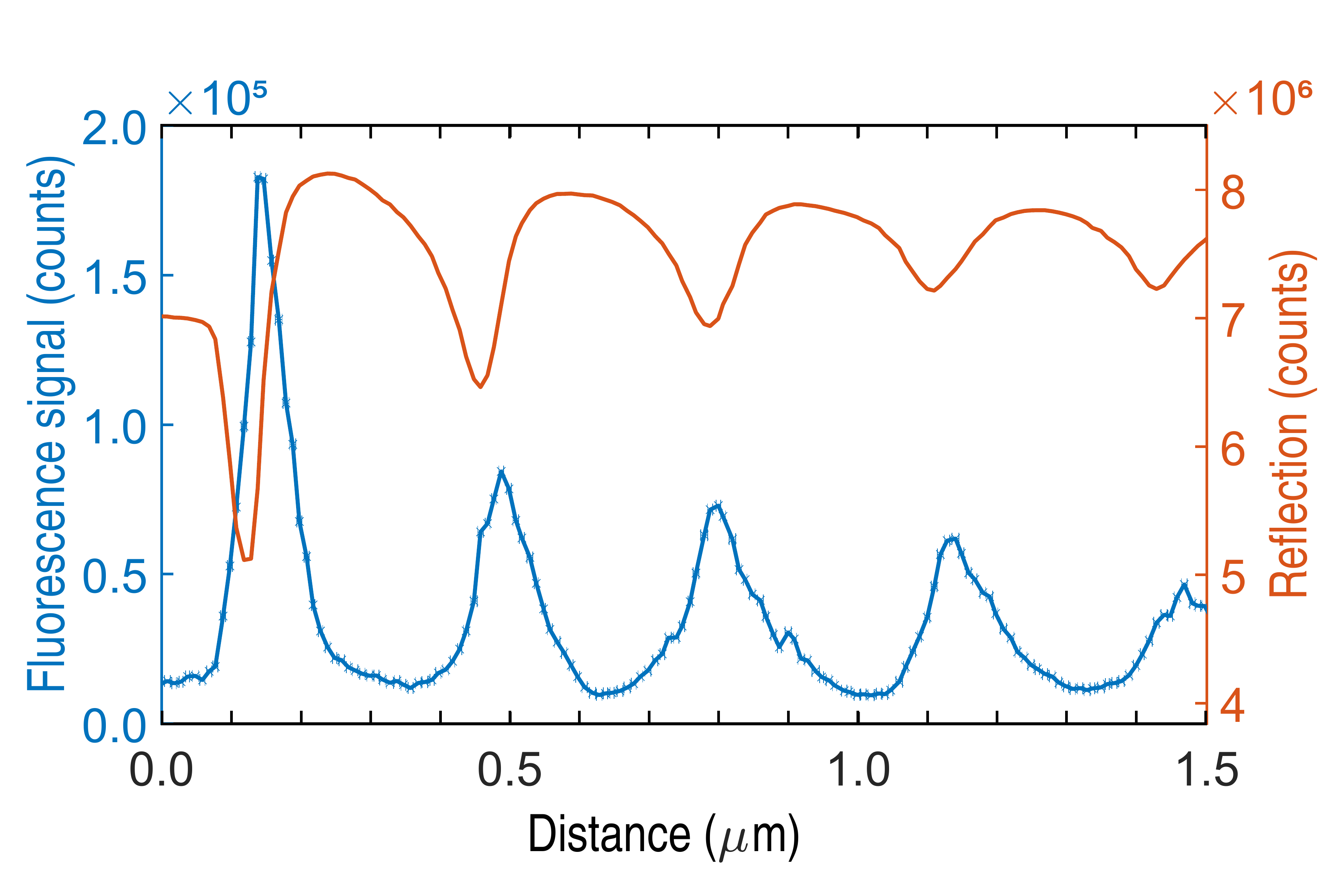}
  \centering
  \caption{\textbf{Fluorescence signal and laser reflection in the planar antenna configuration.} The fluorescence signal consisting of the bead and the director emission (blue curve) as well as the reflected laser power (red curve) are shown as a function of distance between gold wire and SiO$_2$ layer (air gap).  The reflection is associated with  the excitation rate, which modulates the fluorescence signal. }
\label{beads_antenna_fast1}
\end{figure}

Figure~\ref{beads_antenna_fast1} plots the intensity detected by SPAD1 and SPAD2 as a function of distance between the director and the SiO$_2$ layer. We note that the detected intensity from the beads is much higher than $P_\mathrm{coll}$ in Fig.~\ref{Beads_antenna1}b, because in the previous configuration only 5\% of the light was collected with SPAD2, but in this configuration 95\%  was detected by SPAD1.


In Fig.~\ref{beads_antenna_fast1} the detected intensity (blue curve) is the convolution of the  fluorescence of the bead ($P_\mathrm{coll}$) and the background emission from gold islands. This will cause a slight shift (roughly 40~nm) in the position of $P_\mathrm{coll}$ maxima and artificially create a larger distance between the maximum of fluorescence and reflection intensity as compared to theory (Fig.~\ref{theory_emission}). 

The spacing between two minima of the laser reflection (red curve) is roughly 330 nm, which should correspond to the cavity resonance at the excitation light wavelength. This is slightly larger than $\lambda/2$, for $\lambda=636$ nm. The 10~nm difference can be related to the piezo calibration and the system vibrations. 
The reflection intensity oscillations gradually decrease with the air gap. This phenomenon can be simply explained by the fact that while the laser light remains focused on the bead, the reflector is not able to refocus the incident light back into the laser mode. In our theoretical considerations the amplitude of the reflection (R) is constant, since the excitation is a plane wave (Fig.~\ref{theory_emission}b).  

The distance between the first two peaks in $P_\mathrm{coll}$ (blue curve in Fig.~\ref{beads_antenna_fast1}) is different than the next ones. This is probably due to the near-field of the emission at $\lambda = 680$ nm, in combination with the intensity enhancement. Moreover, since the antenna effect has a higher impact at sub-wavelength distances ($\lambda /6 n$ to $ \lambda /4 n$), the first maximum in $P_\mathrm{coll}$ is 2.4 times larger than the second one. 
On the other hand, the first minimum in $R$ is 1.8 times deeper than the second one. By dividing these numbers we get 1.33, which is equal to the first and second peak difference of $P_\mathrm{coll}$ in theory (Fig.~\ref{theory_emission}b).

As the distance increases, $P_\mathrm{coll}$  roughly follows $1-R$, as expected from the predominant contribution of the intensity enhancement. However, in agreement with the simulation results (Fig.~\ref{theory_emission}b), the maxima of $P_\mathrm{coll}$ are slightly ahead of the minima of $R$, because they are related to $|E|^2$. 

\section{Conclusions}

We have proposed and investigated, both theoretically and experimentally, a scanning planar Yagi-Uda antenna, to better understand its influence on the optical properties of nanoscale light sources. Simulations and measurements of the excitation enhancement, the beaming effect, and the Purcell enhancement, give us  insight into the underlying processes and help us maximize fluorescence collection efficiencies over a large parameter range.
These physical quantities are simply addressed by scanning the distance between the antenna elements, the reflector, and the director. Moreover, the experimental findings are consistent and in good agreement with the semi-analytical model.
The results clearly demonstrate that at around 160~nm antenna air gap the  Purcell factor is small, but not negligible (around 2 at the first maximum), and the FWHM of the main radiation lobe at this position is 45$^{\circ}$. All together, the fluorescence signal is 3 fold higher than that on a regular glass coverslip. A comparable improvement can also be found for larger distances.

Compared to other scanning cavity approaches~\cite{colombe2007strong, toninelli2010scanning,kelkar2015sensing}, our method is broadband and less sensitive to the fine position control of the emitters with respect to the antenna elements. Therefore, it is particularly advantageous for detecting fluorescence at ambient temperatures, where the signal is spectrally broad and it allows standard immobilization techniques of the emitters, such as spin-casting or chemical functionalization of the substrate surface. We already envision that we can easily extend this technique to the detection of emitters in a liquid environment. 

Since planar antennas strongly influence the emission directionality, our findings suggest that this approach can be further extended to low-NA optics, particularly to low-NA objectives with long working distances. Avoiding expensive and sometimes impractical high-NA objectives would facilitate the experiments and extend the range of possible applications. Eventually, the tip of an optical fiber coated with a thin gold film (director) could completely replace the objective and as such optical fibers would be employed both for the excitation of the emitters and fluorescence collection~\cite{soltani2019planar}. 
Hence, a scanning planar Yagi-Uda antenna holds promise for a variety of applications in fluorescence-based biosensors~\cite{wolfbeis2005materials}, single-photon sources~\cite{benedikter2017cavity}, enhanced spectroscopy~\cite{gagliardi2014cavity} and scanning microscopy~\cite{shotton1989confocal}.

\subsection*{Funding}
University of Siegen; the German Research Foundation (DFG) (INST 221/118-1 FUGG); and the Bundesministerium f\"{u}r Bildung und Forschung (13N14746).

\subsection*{Acknowledgments}
The authors would like to thank C. Toninelli,  P. Lombardi and G. Shafiee for helpful discussions. Part of this work was performed at the Micro- and Nanoanalytics Facility (MNaF) of the University of Siegen.

\subsection*{Disclosures}
The authors declare no conflicts of interest.

\subsection*{Supplemental document}
See Supplement 1 for supporting content. It includes
experimental setup,  decay rates and deconvolution of the collected signal, collection efficiency with a reflector (semi-antenna) and directionality of planar antenna at larger distances.

\bibliography{sample}

\end{document}